\documentclass[10pt]{emulateapj}
\usepackage{natbib}
\usepackage{color}
\newcommand{\red}[1]{{#1}}

\shorttitle{Validity of the Taylor Hypothesis}
\shortauthors{Howes}

\newcommand\Alfven{Alfv\'en }
\newcommand\Alfvenic{Alfv\'enic }
\newcommand{\V}[1]{\mathbf{#1}} 
 
\newcommand{\zhat}{\mbox{$\hat{\mathbf{z}}$}}

\newcommand{\figref}[1]{Figure~\ref{#1}}
\newcommand{\secref}[1]{\S\ref{#1}}
\newcommand{\eqref}[1]{equation~(\ref{#1})}

\newcommand{\Order}{\mbox{$\mathcal{O}$}} 

\begin{document}


\title{Validity of the Taylor Hypothesis for Linear Kinetic Waves 
in the Weakly Collisional Solar Wind}


\author{G.~G. Howes,  K.~G.~Klein}
\affil{Department of Physics and Astronomy, University of
Iowa, Iowa City, IA, 52242}
\and 
\author{J.~M.~TenBarge }
\affil{IREAP, University of Maryland, College Park, MD 20742}

\begin{abstract}
The interpretation of single-point spacecraft measurements of 
solar wind turbulence is complicated by the fact that the measurements
are made in a frame of reference in relative motion with respect to
the turbulent plasma. The Taylor hypothesis---that temporal
fluctuations measured by a stationary probe in a rapidly flowing fluid
are dominated by the advection of spatial structures in the fluid rest
frame---is often assumed to simplify the analysis.  But measurements
of turbulence in upcoming missions, such as \emph{Solar Probe Plus},
threaten to violate the Taylor hypothesis, either due to slow flow of
the plasma with respect to the spacecraft or to the dispersive nature
of the plasma fluctuations at small scales. Assuming that the
frequency of the turbulent fluctuations is characterized by the
frequency of the linear waves supported by the plasma, we evaluate the
validity of the Taylor hypothesis for the linear kinetic wave modes in
the weakly collisional solar wind. The analysis predicts that a
dissipation range of solar wind turbulence supported by whistler waves
is likely to violate the Taylor hypothesis, while one supported by
kinetic \Alfven waves is not.
\end{abstract}


\section{Introduction}
Developing a thorough understanding of turbulence in space and
astrophysical plasmas is a grand challenge that has the potential to
impact a wide range of research frontiers in plasma physics, space
physics, and astrophysics. The effort to unravel the complex plasma
physical processes that govern the evolution and impact of plasma
turbulence is greatly aided by the ability to make \emph{in situ}
spacecraft measurements of turbulence in the weakly collisional solar
wind plasma. But the interpretation of spacecraft measurements of the
turbulent plasma and electromagnetic field fluctuations is complicated
by the unavoidable fact that the measurements are made in a frame of
reference (the spacecraft frame) that is in relative motion with
respect to the frame of reference of the solar wind plasma (the plasma
frame). 

For each spatial Fourier mode with wavevector $\V{k}$, the
transformation from the frequency $\omega$ in the plasma frame to the
frequency $\omega_{sc}$ in the spacecraft frame yields the relation
$\omega_{sc}= \omega + \V{k}\cdot \V{v}_{sw}$, as shown in
\secref{sec:transform}.  The first term on the right-hand side is the
plasma-frame frequency, and the second term is the advection term
accounting for the frequency arising from the sweeping of a spatial
fluctuation with wavevector $\V{k}$ past the spacecraft at velocity
$\V{v}_{sw}$. Taking advantage of the typically super-\Alfvenic
velocity of the solar wind, $v_{sw}\gg v_A$, observers have
historically adopted the Taylor hypothesis \citep{Taylor:1938},
assuming that $|\omega| \ll |\V{k}\cdot \V{v}_{sw}|$, so that the
spacecraft-frame frequency of fluctuations is interpreted to be
related directly to the wavenumber of the spatial fluctuations in the
plasma frame, $\omega_{sc}\simeq \V{k}\cdot \V{v}_{sw}$
\citep{Matthaeus:1982b,Perri:2010a}.

The validity of using the Taylor hypothesis to transform from
frequency to wavenumber is threatened as upcoming spacecraft missions
push turbulence measurements into new regimes where the Taylor
hypothesis may be violated. Specifically, the Taylor hypothesis may
fail in two distinct regimes: (i) \emph{Slow Flow Regime}: when the
upcoming \emph{Solar Probe Plus} mission samples up to and within the
\Alfven critical point, the solar wind flow velocity will drop to and
fall below the \Alfven velocity, $v_{sw}\lesssim v_A$; and (ii)
\emph{Dispersive Regime}: the high cadence of turbulence measurements
on the upcoming \emph{Magnetospheric Multiscale (MMS)}, \emph{Solar
  Orbiter}, and \emph{Solar Probe Plus} missions effectively samples
spatial fluctuations of ever smaller scale, length scales where the
linear plasma response becomes dispersive, leading to a more rapid
than linear increase of plasma-frame fluctuation frequency with
increasing spatial wavenumber (decreasing length scale). In either of
these cases, the temporal variation of the turbulent fluctuations in
the plasma frame may become non-negligible compared to the temporal
variation due to the sweeping of spatial structure past the
spacecraft, $|\omega| \gtrsim |\V{k}\cdot \V{v}_{sw}|$, thereby
violating the Taylor hypothesis.  The aim of this paper is to explore
the limits of validity of the Taylor hypothesis in the study of
turbulence in the solar wind.

To evaluate the validity of the Taylor hypothesis for solar wind
turbulence, it is necessary to estimate the plasma-frame frequency of
the turbulent fluctuations. A fundamental premise of this study is
that \emph{the frequency of the turbulent fluctuations is well
  characterized by the frequency of the linear waves supported by the
  solar wind plasma}. This concept is contained within a more general
hypothesis for the modeling of plasma turbulence, the
\emph{quasilinear premise} \citep{Klein:2012,Howes:2014b}.  The
quasilinear premise\footnote{Note that the quasilinear premise is not
  the same as \emph{quasilinear theory} in plasma physics, the
  rigorous application of perturbation theory to explore the long-time
  evolution of weakly nonlinear systems.} states simply that
\emph{some} characteristics of turbulent fluctuations in a magnetized
plasma may be usefully modeled by a superposition of randomly-phased,
linear wave modes. The nonlinear interactions inherent to the
turbulent dynamics may be considered to transfer energy among these
linear wave modes---therefore, the model is quasilinear. Here, we
simply adopt the premise that linear wave modes adequately
characterize the frequency response of the turbulent plasma, and we
defer a detailed discussion of the quasilinear premise and supporting
evidence to a subsequent work \citep{Howes:2014b}.  Note that the
weakly collisional nature of the solar wind plasma requires that
kinetic plasma theory, rather than commonly used fluid descriptions
such as magnetohydrodynamics (MHD), must be used to describe the
relevant linear wave modes in the solar wind \citep{Klein:2012}.
Therefore, in this paper, we evaluate the validity of the Taylor
hypothesis for the linear kinetic wave modes in the weakly collisional
solar wind plasma.

\begin{figure*}[t]
\vspace*{2mm}
\begin{center}
\hbox to \hsize{\resizebox{8.3cm}{!}{
\includegraphics*{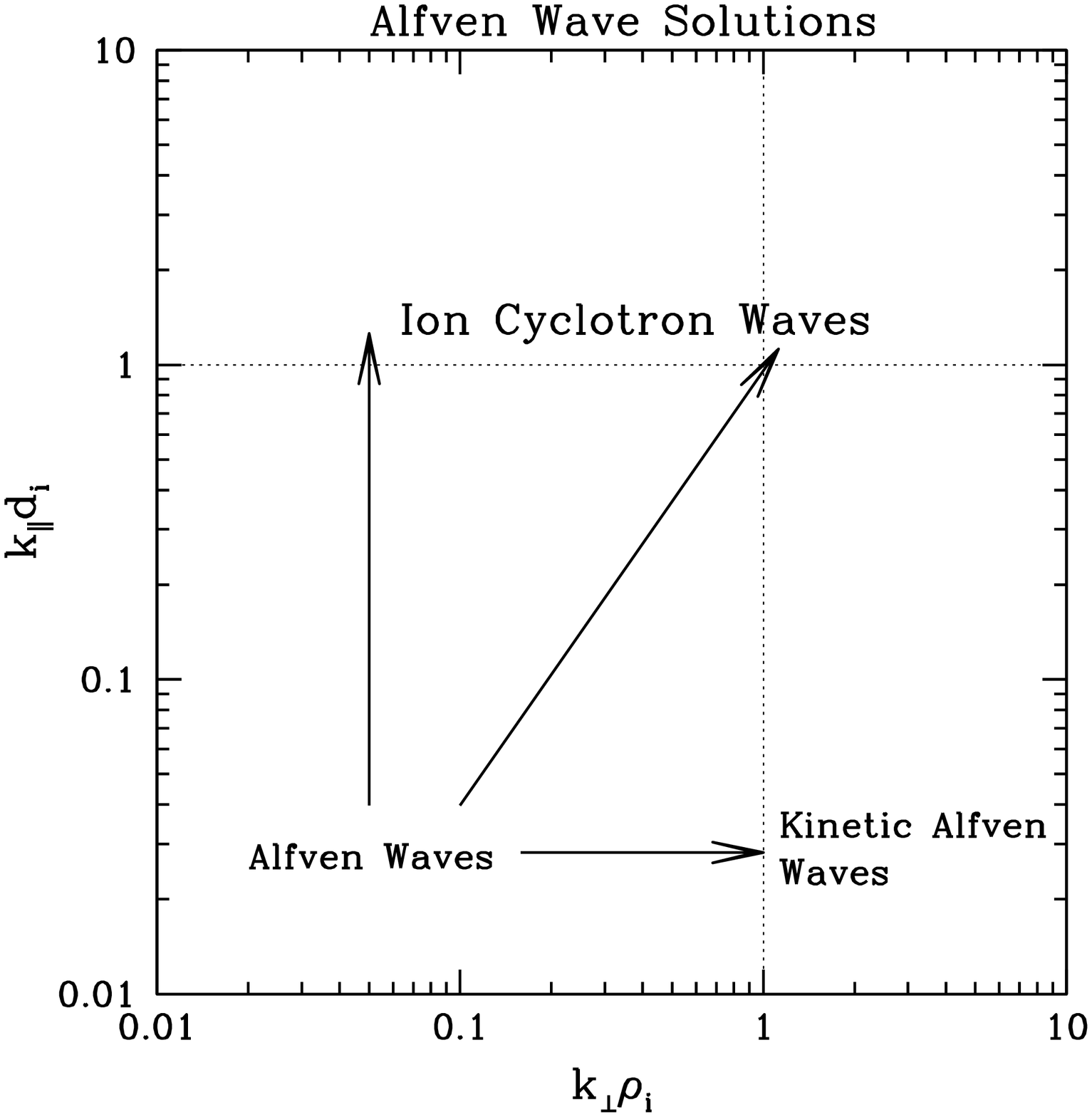}} \hfill
\resizebox{8.3cm}{!}{
\includegraphics*{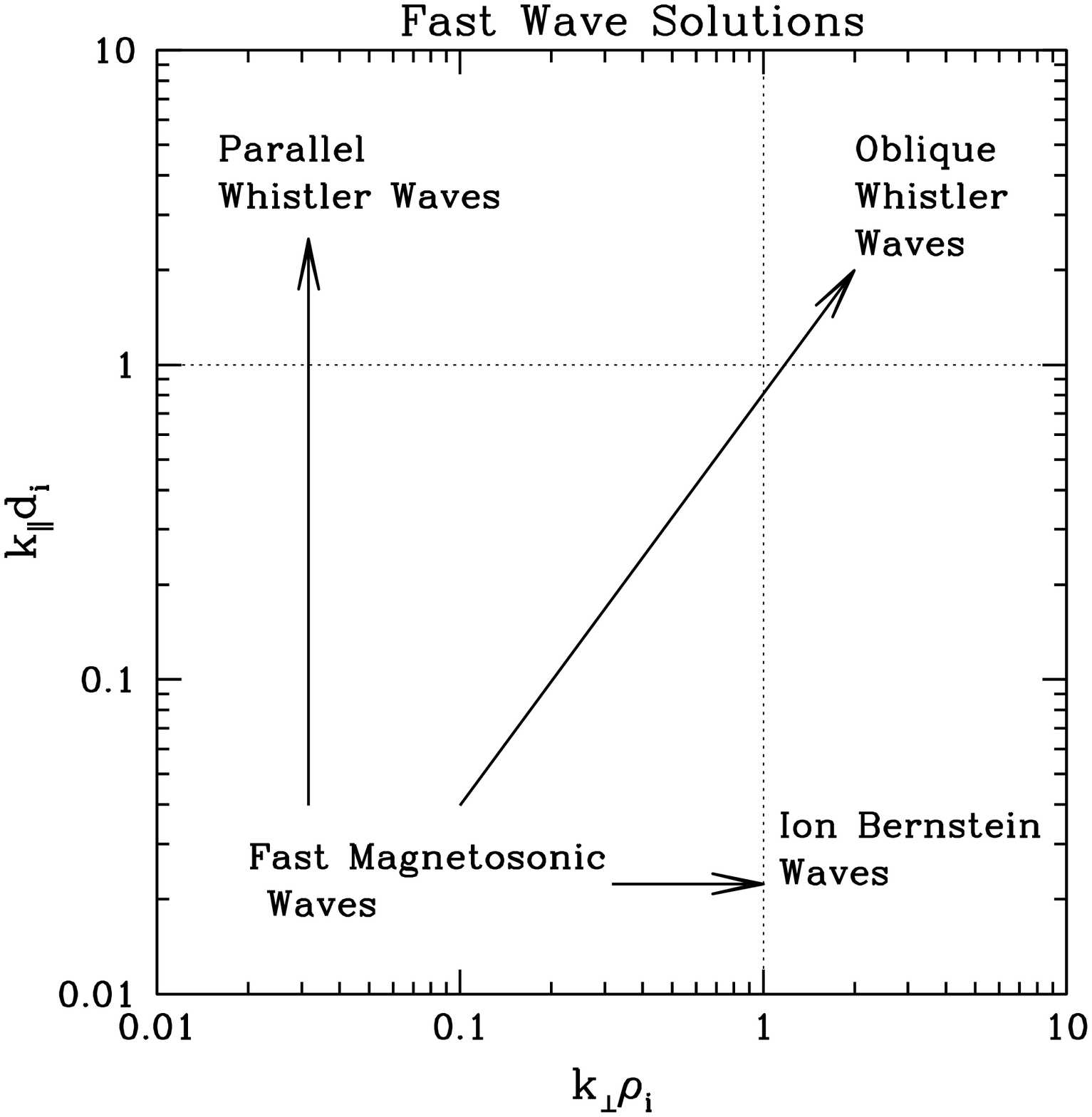}}}
\end{center}
\caption{(a)~Diagram of the region of $(k_\perp,k_\parallel)$ wavevector space
inhabited by the linear wave modes of the \Alfven wave branch for a
collisionless kinetic plasma governed by the Vlasov-Maxwell equations.
(b)~Same for the fast wave branch.}
\label{fig:diagram}
\end{figure*}

The low-frequency wave modes of interest in the study of the inertial
range of solar wind turbulence are the kinetic counterparts of the
fast, Alfv\'en, and slow MHD wave modes \citep{Klein:2012}.  The slow
wave is generally considered to be strongly damped by collisionless
mechanisms for the finite ion temperatures typical of the solar wind
\citep{Barnes:1966}, so previous investigations have generally focused
on the fast and \Alfven wave modes only.  \red{Note, however, that recent
analyses of correlations between different components of the turbulent
fluctuations in the inertial range suggests that the compressible
fluctuations arise from kinetic slow waves
\citep{Smith:2007,Howes:2012a,Klein:2012}, so further investigation of
the role of slow waves in solar wind turbulence is warranted.} But
since the frequency of the slow wave scales in the same way as the
frequency of the \Alfven wave, the conditions for the validity of the
Taylor hypothesis for the slow wave will be similar to that for the
\Alfven wave.  Therefore, we focus here on the remaining two modes,
the \Alfven waves and the kinetic fast waves in the weakly collisional
solar wind plasma at length scales corresponding to both the inertial
range and the dissipation range of solar wind turbulence.

As the turbulent cascade enters the dissipation range, around the
scale of the characteristic ion kinetic length scales, these wave
modes transition to a variety of different dispersive wave modes, with
properties dependent on the region of wavevector space inhabited by
each mode. \figref{fig:diagram} depicts these modes, and the regions
of wavevector space in which they exist, for the \Alfven and fast wave
branches.  Since the ion cyclotron waves of the \Alfven branch are
strongly damped by cyclotron resonance with the ions and ion Bernstein
waves of the fast wave branch are dominantly electrostatic in
nature,\footnote{Although, see \citet{Sahraoui:2012} for a more
  in-depth exploration of the magnetic signature of ion Bernstein
  wave.}  the electromagnetic fluctuations observed in the solar wind
dissipation range are believed to be one, or a mixture, of the two
remaining linear kinetic wave modes: whistler waves and kinetic
\Alfven waves \citep{Howes:2009b}.

Here we derive simple analytical expressions for the validity of the
Taylor hypothesis for the study of plasma turbulence in the solar
wind, focusing on the turbulence at scales smaller than the ion length
scales, consisting of whistler waves or kinetic \Alfven waves. In
\secref{sec:th}, we derive a useful form of the Taylor hypothesis and
derive the general condition for its validity. In \secref{sec:valid},
we derive simplified analytical expressions for the conditions
necessary for the Taylor hypothesis to be valid for MHD \Alfven waves,
kinetic \Alfven waves, ion cyclotron waves, kinetic fast MHD waves,
and whistler waves. These expressions use simple approximate
dispersion relations for the \Alfven and kinetic \Alfven waves and for
the whistler waves that are validated against numerical solutions of
the linear Vlasov-Maxwell dispersion relation in \secref{sec:num}. We
discuss the consequences of these findings in \secref{sec:discuss} for
the study of both the inertial range and the dissipation range of
solar wind turbulence. Finally, we conclude in \secref{sec:conc} with
the primary prediction of this paper that a dissipation range of 
turbulence supported by whistler waves will violate the Taylor
hypothesis, while a dissipation range consisting of kinetic \Alfven
waves typically will not.  The qualitative and quantitative effects of
the violation of the Taylor hypothesis on the magnetic energy spectrum
in the solar wind that will be measured by upcoming missions, such as
\emph{Solar Probe Plus}, are discussed in a companion paper,
\citet{Klein:2014b}.

\section{The Taylor Hypothesis}
\label{sec:th}
In this section, we derive the relationship between the plasma-frame
and spacecraft-frame frequency, express the validity condition for the
Taylor hypothesis in a convenient normalization, and discuss the
conditions under which the Taylor hypothesis may be violated.

\subsection{Transforming from Plasma-Frame to Spacecraft-Frame Frequency}
\label{sec:transform}
The magnetic field in the solar wind plasma frame as a function of
position and time, $\V{B}(\V{r},t)$, may be expressed in general as a
sum of Fourier components in wavevector $\V{k}$ and frequency $\omega$
by
\begin{equation}
\V{B}(\V{r},t)  =  \sum_{\V{k}}\sum_\omega \hat{\V{B}}(\V{k},\omega) e^{i(\V{k}\cdot \V{r} - \omega  t)}.
\end{equation} 
Note that, in general, each wavevector $\V{k}$ can have separate
contributions at different frequencies, $\omega$.

We sample this solar wind magnetic field at a spacecraft moving at
velocity $-\V{v}_{sw}$, equivalent to spacecraft measurements where the
solar wind is streaming past a stationary spacecraft at velocity
$\V{v}_{sw}$. The position of the spacecraft as a function of time is
given by $\V{r}=-\V{v}_{sw}t$, so the time series of magnetic field
measurements in the spacecraft frame is given by
$\V{B}(t)=\left. \V{B}(\V{r},t)\right|_{\V{r}=-\V{v}_{sw}t}$, yielding
\begin{equation}
\V{B}(t) = \sum_{\V{k}}\sum_\omega  \hat{\V{B}}(\V{k},\omega)
e^{-i[\V{k}\cdot \V{v}_{sw} + \omega] t}.
\end{equation} 
Finally, we Fourier transform the spacecraft-frame magnetic field time
series to obtain the signal in terms of the spacecraft-frame frequency
$\omega_{sc}$ given by $\V{B}(\omega_{sc}) = (1/2 \pi)\int dt \  \V{B}(t)
e^{i \omega_{sc} t}$, or
\begin{equation}
\V{B}(\omega_{sc}) =  \sum_{\V{k}} \sum_\omega \hat{\V{B}}(\V{k},\omega)
\delta[\omega_{sc}-\V{k}\cdot \V{v}_{sw} - \omega] .
\end{equation} 

Thus we find that the spacecraft-frame frequency is given by the
argument of the delta function, 
\begin{equation}
\omega_{sc}=  \omega + \V{k}\cdot \V{v}_{sw}.
\label{eq:doppler}
\end{equation} 
Note that this condition must be applied separately to each Fourier
mode $\V{k}$.  \emph{A serious limitation of single-point spacecraft
  measurements is that it is not, in general, possible to separate
  these two contributions to the measured spacecraft-frame frequency,
  $\omega_{sc}$}.

\subsection{Applying the Quasilinear Premise}
\label{sec:qlp}
For all possible linear wave modes $m$ with frequency
$\omega=\omega_m(\V{k})$, we may apply the premise that the frequency
of the turbulent fluctuations is well characterized by the frequency
of the linear waves by summing over all possible linear wave modes $m$ and
multiplying by the constraint $\delta[\omega-\omega_m(\V{k})]$ to obtain
\begin{equation}
\V{B}(\omega_{sc}) =  \sum_m \sum_{\V{k}} \hat{\V{B}}_m(\V{k})
\delta[\omega_{sc}-\V{k}\cdot \V{v}_{sw} -  \omega_m(\V{k})] .
\end{equation} 
To simplify the notation, we have defined
$\hat{\V{B}}_m(\V{k})\equiv\hat{\V{B}}[\V{k},\omega_m(\V{k})]$ since,
according to our premise, the plasma-frame frequency is completely
determined by the wave mode $m$ and wavevector $\V{k}$.  With the
implicit understanding that the plasma-frame frequency $\omega$ is
constrained to be given by the linear dispersion relation for mode
$m$, $\omega=\omega_m(\V{k})$, the relation between the spacecraft
frame frequency and the plasma-frame frequency is the same as that
given in \eqref{eq:doppler}.

\subsection{Validity Condition for the Taylor Hypothesis}

We can better understand the relative contributions to $\omega_{sc}$
of the two terms on the right-hand side of \eqref{eq:doppler} if we
normalize the equation by a characteristic frequency for waves in a
magnetized plasma, $k v_A$, yielding
\begin{equation}
\frac{\omega_{sc}}{k v_A}=\frac{\omega }{k v_A}+ \frac{v_{sw}}{v_A} \cos \theta,
\label{eq:eqnorm}
\end{equation}
where we have specified a wavevector $\V{k}$ at an angle $\theta$ with
respect to the solar wind velocity, such that $\V{k}\cdot\V{v}_{sw}= k
v_{sw} \cos \theta$. 

Single-point spacecraft measurements are often analyzed by taking
advantage of the fact that the solar wind flows past the spacecraft at
generally super-\Alfvenic velocities, $v_{sw}\gg v_A$, where a typical
velocity ratio in the solar wind is $v_{sw}/v_A\simeq 10$
\citep{Tu:1995,Bruno:2005}.  In contrast, the first term on the
right-hand side of \eqref{eq:eqnorm} has a typical magnitude 
$\omega/(k v_A) \lesssim 1$, so this first term is often negligible.
For a system in which temporal fluctuations are measured by a
stationary probe in a rapidly flowing fluid, the Taylor hypothesis
makes the approximation that the frequency measured by the probe is
dominated by the advection of spatial fluctuations past the probe,
$|\omega| \ll |\V{k} \cdot \V{v}_{sw}|$ \citep{Taylor:1938}. In the case
of the solar wind, this means taking $\omega_{sc} \simeq \V{k}\cdot
\V{v}_{sw}$. Therefore, the Taylor hypothesis is valid in the limit
\begin{equation}
\frac{v_{sw}}{v_A} \cos \theta \gg \frac{\omega }{k v_A}.
\label{eq:valid}
\end{equation}
It is important to point out that this condition must be evaluated for
each wavevector $\V{k}$ and wave mode $m$ that makes up the turbulent
distribution of power in three-dimensional wavevector space.

\subsection{Violation of  the Taylor Hypothesis}
Violation occurs when $|\omega| \gtrsim |\V{k} \cdot \V{v}_{sw}|$,
which can occur in the \emph{Slow Flow Regime} or the \emph{Dispersive
  Regime}. In the Slow Flow Regime, nondispersive waves may violate
the Taylor hypothesis when the wave velocity is of order or greater
than the flow velocity. In the Dispersive Regime, dispersive effects
causing the wave frequency to increase more rapidly than linearly with
the wavevector\footnote{\red{For resonances in the dispersive regime,
    the Taylor hypothesis becomes increasingly well satisfied at
    smaller scales since the wave frequency $\omega \rightarrow$
    constant as $k \rightarrow \infty$, so the plasma-frame term in
    \eqref{eq:doppler} remains constant while the advection term
    increases linearly with $k$.  This is shown quantitatively for ion
    cyclotron waves in \secref{sec:icw}.}} may lead to high-frequency
turbulent fluctuations that violate the Taylor hypothesis.

Since turbulence theories typically predict that turbulent power fills
a region of three-dimensional wavevector space, it is generally
possible that turbulent power exists in wavevectors $\V{k}$ that are
perpendicular to the solar wind velocity, $\theta \rightarrow \pi/2$.
In this case, the factor $\cos \theta \rightarrow 0$, and therefore
the validity condition for the Taylor hypothesis \eqref{eq:valid} is
not satisfied (unless the plasma-frame frequency $\omega=0$,
\emph{i.e.}~a convected structure that does not evolve in time in the
plasma frame, such as a pressure-balanced structure or an entropy mode
fluctuation). In fact, for all wavevectors in the volume of wavevector
space with angles such that $\cos \theta \ll 1$, the Taylor hypothesis
will be violated.  But, this volume of wavevectors that violate
\eqref{eq:valid} is often vanishingly small.

For example, consider the case of turbulence that fills wavevector
space isotropically, such as is predicted for hydrodynamic turbulence
or the fast wave component of MHD turbulence \citep{Cho:2003}.  In this
case, the turbulent amplitudes vary with spherical radius but do not
depend on the azimuthal or polar angles in spherical coordinates. For
the purpose of this illustrative example, let us assume that $\omega
/(kv_A) \sim 1$, so the validity condition \eqref{eq:valid} will be
significantly violated when $ \cos \theta \le v_A/v_{sw}$. For a
typical ratio of $v_{sw}/v_A=10$, this significant violation will
occur only for angles $\theta>\theta_c = \cos^{-1}
(v_A/v_{sw})=84^\circ$. The ratio of the volume of wavevector space
which significantly violates the Taylor hypothesis to the total volume
of wavevector space is given by $\cos \theta_c = v_A/v_{sw}= 0.1$, so
only 10\% of the turbulent power violates the Taylor hypothesis in
this example.

The case for magnetized plasma turbulence is significantly more
complicated because the turbulent power is predicted theoretically
\citep{Goldreich:1995,Boldyrev:2006,Howes:2008b,Howes:2008c,Schekochihin:2009},
shown numerically
\citep{Shebalin:1983,Cho:2000,Maron:2001,Cho:2004,Saito:2008,Gary:2012,TenBarge:2012a},
and measured observationally
\citep{Sahraoui:2010b,Narita:2011,Roberts:2013} to fill wavevector
space anisotropically, with most power concentrated in wavevectors
nearly perpendicular to the local mean magnetic field.  Note that the
angle of the wavevector with respect to the local mean magnetic field
is \emph{not} the same as the angle $\theta$ of the wavevector with
respect to the solar wind velocity.  \red{However, in the near-sun
  enivronment that will be probed by \emph{Solar Orbiter} and
  \emph{Solar Probe Plus}, the Parker spiral magnetic field is nearly
  radial, so in this region the long-time-averaged magnetic field is
  indeed more often aligned with the solar wind plasma flow.  In this
  case, wavevectors nearly perpendicular to the averaged magnetic
  field will also have $\theta \rightarrow \pi/2$. Note, however, that
  at the outer scale of the turbulent inertial range (based on
  spacecraft measurements at heliocentric distances of $0.3$~AU and
  greater), the magnetic field fluctuations have $|\delta \V{B}| \sim
  |\V{B}_0|$, so although the average magnetic field may be aligned
  with the solar wind flow, instantaneously the \emph{local} magnetic
  field direction may have a wide range of angles about the Parker
  spiral value. In addition, the} anisotropic distribution of
turbulent wave power also leads to lower plasma-frame frequencies for
\Alfvenic fluctuations: since $k_\parallel \ll k_\perp$ and $\omega =
k_\parallel v_A$, this means that $\omega /(kv_A) \ll 1$, leading to a
much narrower volume of wavevector space in which the Taylor
hypothesis may be significantly violated.  Therefore, one needs a more
sophisticated approach to evaluate the validity of the Taylor
hypothesis for anisotropic plasma turbulence.  In our companion work,
\citet{Klein:2014b}, we employ the synthetic spacecraft data method
\citep{Klein:2012} both to investigate the conditions for the validity
of the Taylor hypothesis and to predict the quantitative effect on the
measured magnetic energy spectrum in spacecraft-frame frequency when
the Taylor hypothesis is violated.

\section{Validity of the Taylor Hypothesis for Kinetic Wave Modes}
\label{sec:valid}
In this section, we will use the properties of the linear wave modes
in a weakly collisional plasma to evaluate the validity of the Taylor
hypothesis. The goal is to derive simple expressions for the
observational identification of conditions in which a particular wave
mode may violate the Taylor hypothesis. We focus on the kinetic
counterparts of the \Alfven and fast waves, at length scales both
above and below the characteristic ion length scales: (1) the thermal
ion Larmor radius, $\rho_i = v_{ti}/\Omega_i$, where the ion thermal
velocity is defined by $v_{ti}^2=2 T_i/m_i$ (absorbing Boltzmann's
constant to measure temperature in energy units) and $\Omega_i=
q_iB_0/(m_ic)$ is the ion cyclotron frequency; and (2) the ion
inertial length, $d_i = v_A/\Omega_i$, where the \Alfven velocity in a
magnetic field with equilibrium magnitude $B_0$ is given by $v_A^2=
B_0^2/(4 \pi n_i m_i)$. The relation between the ion Larmor radius and
ion inertial length is given by $\rho_i=d_i\sqrt{\beta_i}$, where the
ion plasma beta is $\beta_i= 8 \pi n_i T_i/B_0^2$.  Note that,
throughout this paper, $k_\parallel$ and $k_\perp$ are defined as the
parallel and perpendicular components of the wavevector with respect
to the direction of the \emph{local} mean magnetic field $\V{B}_0$.

\red{We emphasize here that we use analytical approximations for the
  linear wave mode frequencies that are either rigorous limits of
  kinetic theory \citep{Howes:2006,Schekochihin:2009} or empirical
  expressions based on the numerical results from the linear
  Vlasov-Maxwell dispersion relation \citep{Quataert:1998,Howes:2006},
  as verified in \secref{sec:num}. The use of analytical expressions
  from fluid theories, such as Hall MHD \citep{Hirose:2004,Ito:2004}
  or two-fluid theory \citep{Stringer:1963}, can lead to a mixed up
  identification of wave modes from the fluid theory with those
  arising from the more broadly applicable kinetic theory. As an
  example, as shown in Figure~3 of \citet{Howes:2009b}, for the finite
  ion temperature conditions relevant to the study of the solar wind,
  Hall MHD (as well as two-fluid theory \citep{Stringer:1963})
  connects the slow wave to the ion cyclotron resonance as
  $k_\parallel d_i \rightarrow 1$, whereas the Vlasov-Maxwell results
  show that this resonance is actually associated with the \Alfven
  wave. Such occasional mixing up of the identification of wave modes
  between fluid theory and kinetic theory has been noted in the
  literature \citep{Krauss-Varban:1994,Yoon:2008,Howes:2009b}, but
  this fact is not necessarily widely known. The bottom line is that,
  to ensure the correct identification of wave modes using fluid
  theory, one should always confirm the results using kinetic theory
  for the appropriate plasma parameters.  All simple analytical
  expressions used in this study originate from the results of kinetic
  theory, and we caution the reader that conflicting information about
  wave mode properties from fluid theories should be carefully checked
  against kinetic results.}

\subsection{\Alfven Modes}
At parallel length scales $k_\parallel d_i \ll 1$, the \Alfven wave
frequency is well estimated by
\begin{equation}
\omega=k_\parallel v_A \sqrt{1 + 
\frac{(k_\perp \rho_i)^2}{\beta_i + 2/(1+T_e/T_i)}},
\label{eq:eq1}
\end{equation}
\citep{Howes:2006}. We numerically verify this dispersion relation in
\secref{sec:num}.  Note that in the regime $k_\parallel d_i \gtrsim
1$, the \Alfven mode becomes the heavily damped ion cyclotron wave
with frequency $\omega \simeq \Omega_i$ (see \figref{fig:diagram});
this limit is discussed separately in \secref{sec:icw}.  Substituting
this frequency into the validity condition given by \eqref{eq:valid},
we obtain
\begin{equation}
\frac{v_{sw}}{v_A} \cos \theta \gg \frac{k_\parallel }{k}\sqrt{1 + 
\frac{(k_\perp \rho_i)^2}{\beta_i + 2/(1+T_e/T_i)}}.
\label{eq:alf1}
\end{equation}
We determine below simplified versions of this validity condition for
limits of  $k_\perp \rho_i$.

\subsubsection{\Alfven Wave Limit, $k_\perp \rho_i \ll 1$}
In the limit $k_\perp \rho_i \ll 1$, the \Alfven mode is the
collisionless counterpart of the usual MHD \Alfven
wave\footnote{Because the \Alfven wave is incompressible, kinetic
effects do not modify the linear wave properties in the limit
$k_\parallel d_i \ll 1$; in fact, it has been proven that the
reduced MHD description of \Alfven waves is a rigorous limit of the
kinetic theory for anisotropic fluctuations with $k_\parallel \ll
k_\perp$ \citep{Schekochihin:2009}.} with frequency
$\omega=k_\parallel v_A$. Therefore, the validity condition for the
\Alfven waves of the solar wind inertial range simplifies to
\begin{equation}
\frac{v_{sw}}{v_A} \cos \theta \gg \frac{k_\parallel }{k}.
\label{eq:alf_mhd}
\end{equation}
In addition to the fact that it is always true that $k_\parallel /k
\le 1$, the turbulence in magnetized plasmas is widely observed to
cascade anisotropically to smaller scales perpendicular than parallel
to the magnetic field, leading to the condition $k_\perp \gg
k_\parallel$
\citep{Shebalin:1983,Cho:2000,Maron:2001,Cho:2004,Sahraoui:2010b,Narita:2011,TenBarge:2012a,Roberts:2013}. In
this case, $k_\parallel /k \simeq k_\parallel /k_\perp \ll 1$. Since
the super-\Alfvenic conditions of the solar wind typically satisfy the
relation $v_{sw}/v_A \gg 1$, we predict that the \Alfvenic
fluctuations in the inertial range of solar wind turbulence do not
typically violate the Taylor hypothesis.

\subsubsection{Kinetic \Alfven Wave Limit, $k_\perp \rho_i \gg 1$}
In the asymptotic range of kinetic \Alfven waves, given by $k_\perp
\rho_i \gg 1$, the validity condition becomes
\begin{equation}
\frac{v_{sw}}{v_A} \cos \theta \gg \frac{k_\parallel }{k}
\frac{k_\perp \rho_i}{\sqrt{ \beta_i + 2/(1+T_e/T_i)}}.
\label{eq:alf_kaw}
\end{equation}
The dispersive nature of kinetic \Alfven waves gives rise to the
factor of $k_\perp \rho_i$ in this expression and may lead to a
violation of the Taylor hypothesis at sufficiently high perpendicular
wavenumber. Because kinetic \Alfven waves have $k_\parallel \ll
k_\perp$, we may replace $k\simeq k_\perp$ in the denominator. The
term $ 2/(1+T_e/T_i)$ simplifies to the value 0 for $T_i/T_e \ll 1$,
to the value 1 for $T_i/T_e =1$, and to the value 2 for $ T_i/T_e \gg
1$, so we take the $T_i/T_e =1$ result as the case most representative
of solar wind conditions. For limits of the ion plasma beta $\beta_i$,
we find the following approximate results for the validity condition
\begin{equation}
\frac{v_{sw}}{v_A} \cos \theta \gg
\left\{\begin{array}{cc}
k_\parallel  \rho_i, & \beta_i \lesssim 1 \\
k_\parallel  d_i, & \beta_i \gtrsim 1 
\end{array}\right. .
\label{eq:alf_kaw_betai}
\end{equation}
Note that at $\beta_i \le 1$, $\rho_i \le d_i$, and so a conservative
constraint at any value of $\beta_i$ may be written as $v_{sw}/v_A \cos
\theta \gg k_\parallel d_i$.  Therefore,  the condition
for kinetic \Alfven waves to satisfy the Taylor hypothesis simplifies
to
\begin{equation}
\frac{v_{sw}}{v_A} \cos \theta \gg  k_\parallel d_i.
\label{eq:alf_kaw_fin}
\end{equation}
Since the kinetic \Alfven wave exists only\footnote{Note, however,
  that at $\beta_i>1$ and $k_\perp\rho_i \gg 1$, the kinetic \Alfven
  wave becomes insensitive to the ion cyclotron resonance and can
  persist at values $k_\parallel d_i >1$ \citep{Boldyrev:2013}.}  in the
wavevector regime $k_\parallel d_i \ll 1$, we predict that the kinetic
\Alfven wave fluctuations in the dissipation range of solar wind
turbulence typically do not violate the Taylor hypothesis.

\subsubsection{Ion Cyclotron Wave Limit, $k_\parallel d_i \gtrsim 1$}
\label{sec:icw}
In the limit $k_\parallel d_i \gtrsim 1$, the \Alfven mode transitions to
the ion cyclotron wave (see \figref{fig:kaw}) and generally suffers
very strong ion cyclotron damping.  Although the expectation is that
the ion cyclotron waves are so strongly damped that they will rarely
be observed in the solar wind, we nevertheless estimate the validity
of the Taylor hypothesis for these waves. Based on solutions of the
Vlasov-Maxwell dispersion relation, we have constructed a simple,
rough estimate of the maximum frequency of ion cyclotron waves
\begin{equation}
\omega \lesssim \frac{k_\parallel d_i + (k_\parallel d_i)^2}{1 + (k_\parallel d_i)^2}\Omega_i,
\label{eq:icw}
\end{equation}
valid for $k_\perp \rho_i \lesssim 1$.  In the \Alfven wave limit,
$k_\parallel d_i \ll 1$, this function simplifies to
$\omega=k_\parallel v_A$, as expected. In the ion cyclotron wave limit
$k_\parallel d_i \gtrsim 1$, this function asymptotes to the ion
cyclotron frequency, $\omega \simeq \Omega_i$.  Although strong
collisionless damping via the ion cyclotron resonance often leads to
Vlasov-Maxwell solutions of the wave frequency that are somewhat below
$\Omega_i$ (by a factor of order unity), \eqref{eq:icw} provides an
upper bound to the frequency and is therefore suitable for exploring
the validity of the Taylor hypothesis.

Substituting \eqref{eq:icw} into \eqref{eq:valid}, we find that the
condition for ion cyclotron waves to satisfy the Taylor hypothesis
becomes
\begin{equation}
\frac{v_{sw}}{v_A} \cos \theta \gg \frac{k_\parallel }{k}
\frac{1 + k_\parallel d_i}{1 + (k_\parallel d_i)^2}.
\label{eq:icw_tay}
\end{equation}
Since $k_\parallel/k \le 1$ and the remaining factor gives a value of
order unity or less for any value of $k_\parallel d_i$, we predict
that the ion cyclotron wave fluctuations in solar wind turbulence
typically do not violate the Taylor hypothesis.

\subsection{Fast Modes}
In this section, we evaluate the validity condition of the Taylor
hypothesis for the kinetic counterpart of the fast MHD wave
\citep{Klein:2012} and for the whistler wave. Because ion Bernstein
waves, the solution of the fast wave branch in the regime $k_\parallel
d_i \ll 1$ and $k_\perp \rho_i \gtrsim 1$ as shown in
\figref{fig:diagram}, are essentially electrostatic in nature
\citep{Stix:1992}, it has been suggested that they are unlikely to be
responsible for the magnetic field fluctuations observed in the
dissipation range of solar wind turbulence \citep{Howes:2009b}. It has
been pointed out, however, that at $k\rho_i \sim 1$, the ion Bernstein
wave magnetic signature is small but nonzero
\citep{Sahraoui:2012}. The possibility that ion Bernstein waves
contribute significantly to solar wind turbulence merits further
consideration, but we leave that to future work and do not consider
ion Bernstein waves here.
\subsubsection{Fast Wave Limit, $k d_i \ll 1$}
In the large scale limit $k d_i \ll 1$, the fast wave mode is the
kinetic counterpart of the usual MHD fast wave \citep{Klein:2012}.  An
upper limit on the frequency of this wave can be expressed as
\begin{equation}
\omega \le   k \sqrt{v_A^2 + c_s^2} \simeq kv_A \sqrt{1 + \beta_i(1+T_e/T_i)},
\label{eq:fast}
\end{equation}
so the validity condition becomes
\begin{equation}
\frac{v_{sw}}{v_A} \cos \theta \gg \sqrt{1 + \beta_i(1+T_e/T_i)}.
\end{equation}
Observed values of $\beta_i$ in the near-Earth solar wind
\citep{Howes:2011c} generally satisfy $\beta_i <5$, so although this
condition as not quite as well satisfied as for the \Alfven modes, we
predict that fast waves do not typically violate the Taylor hypothesis
in a significant way. Note that, since the kinetic fast wave at $k d_i
\ll 1$ is nondispersive, any violation of the Taylor hypothesis is
necessarily due to plasma conditions in the Slow Flow Regime.

\subsubsection{Whistler Wave Limit, $k_\parallel d_i \gtrsim 1$}
Whistler waves are the manifestation of the kinetic fast mode in the
limit $k_\parallel d_i \gtrsim 1$, and their frequency is well
estimated by
\begin{equation}
\omega =  k v_A \sqrt{ 1+ (k_\parallel  d_i)^2}.
\label{eq:whistler}
\end{equation}
We verify numerically in \secref{sec:num} that this simplified
dispersion relation provides a good estimate of the whistler wave
frequency.  Any violation of the Taylor hypothesis is likely to happen
in the regime $k_\parallel d_i \gg 1$, leading to a validity condition
\begin{equation}
\frac{v_{sw}}{v_A} \cos \theta \gg k_\parallel d_i.
\end{equation}
Here we see that if a whistler wave has a sufficiently high value of
$k_\parallel d_i$, it will violate the Taylor hypothesis.  This
violation occurs in the Dispersive Regime. Therefore, given measured
values for the \Alfven velocity $v_A$ and solar wind velocity
$v_{sw}$, it is possible to determine the value of the parallel
wavevector at which the whistler wave will violate the Taylor
hypothesis.

Single-point spacecraft measurements, however, do not allow the
determination of the parallel and perpendicular components of the
wavevector with respect to the \emph{local mean magnetic
  field}. Therefore, we choose to express the condition in terms of an
observable quantity, the effective projection of the wavevector
$k_{\small{\mbox{eff}}} \simeq k \cos \theta$ along the direction of the solar wind
flow.\footnote{Strictly, this is true only when the Taylor hypothesis
  \emph{is} valid.} Since whistlers have wavevectors that satisfy
$k_\parallel \gtrsim k_\perp$, we first make the simplification
$k_\parallel \simeq k$. Then, we approximate the magnitude of the
wavevector by its projection along the solar wind flow, $k_{\small{\mbox{eff}}} \simeq k
\cos \theta$. The resulting condition for the validity of the Taylor
hypothesis, in terms of the measured component of the wavevector along
the flow direction $k_{\small{\mbox{eff}}}$, is
\begin{equation}
\frac{v_{sw}}{v_A} \cos^2 \theta \gg k_{\small{\mbox{eff}}} d_i,
\end{equation}
where $ k_{\small{\mbox{eff}}}\simeq \omega_{sc}/v_{sw}$ is the observable
component of the wavevector sampled along the direction of the solar
wind flow.

Taking $\cos \theta \sim 1$, we can express the spacecraft-frame
frequency at which we expect the Taylor hypothesis to be violated by
\begin{equation}
\omega_{sc} \gtrsim \frac{v_{sw}^2}{v_A d_i}.
\end{equation}
In the notation used in \citet{Klein:2014b}, $\overline{V} \equiv
v_{sw}/v_A$ and $\omega_*/\Omega_i\equiv
(\omega_{sc}/\Omega_i)/\overline{V}$, this condition becomes
\begin{equation}
\frac{\omega_*}{\Omega_i} \gtrsim \overline{V}.
\end{equation}

\subsection{Slow Modes}
The kinetic counterpart of the MHD slow mode is strongly damped in
collisionless plasmas with $T_i\sim T_e$
\citep{Barnes:1966,Klein:2012}, so previous analyses have generally
dismissed the possibility of kinetic slow waves in the solar wind.
Recently, however, the first study to employ Vlasov-Maxwell kinetic
theory to investigate the properties of the compressible fluctuations
in the inertial range of the solar wind has found strong observational
evidence that the compressible fluctuations are almost entirely in the
kinetic slow mode \citep{Howes:2012a,Klein:2012}.

The normalized phase velocity of the kinetic slow wave $\omega/(k
v_A)$ is almost always less than the kinetic fast wave---see Figure~1
of \citet{Klein:2012}---and has properties generally similar to the
\Alfven wave.  Since the frequency of the slow mode scales in the same
way as the \Alfven mode, we do not separately address the slow mode
but use the results for the \Alfven mode as a guide.  Therefore, based
on our prediction that the \Alfven and fast wave modes do not violate
the Taylor hypothesis, we predict that the kinetic slow mode will also
not violate the Taylor hypothesis.

\section{Numerical Verification of Linear Wave Frequency Estimates}
\label{sec:num}
In this section, we verify that the simple expressions for the linear
frequencies of the Alfv\'en, kinetic Alfv\'en, and whistler waves
provide good estimates of the frequencies given by the linear
Vlasov-Maxwell dispersion relation \citep{Quataert:1998,Howes:2006}. A
fully-ionized proton and electron plasma is assumed with isotropic
Maxwellian velocity distributions, a realistic mass ratio
$m_i/m_e=1836$, and non-relativistic conditions $v_{ti}/c = 10^{-4}$.
For the comparison presented here, we choose plasma parameters that
are typical of near-Earth solar wind conditions, $ \beta_i=1$ and
$T_i/T_e=1$.  The complex linear Vlasov-Maxwell eigenfrequencies for
the \Alfven mode and fast mode are solved on a logarithmically spaced
grid in wavevector space spanning $10^{-2} \le k_\parallel d_i \le 10$
and $10^{-2} \le k_\perp d_i \le 10$.  Note that for a $\beta_i=1$
plasma, the thermal ion Larmor radius and ion inertial length are the
same, $\rho_i=d_i$.

\begin{figure}[t]
\resizebox{3.1in}{!}{\includegraphics{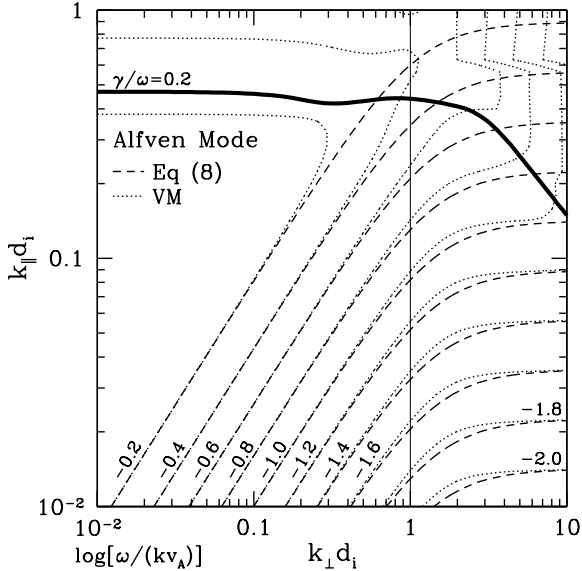}}
\caption{Contour plot of  $\log[\omega/(k v_A)]$ for the \Alfven mode over the 
$(k_\perp d_i, k_\parallel d_i)$ plane for a $\beta_i=1$ and $T_i/T_e=1$ plasma determined by the linear
Vlasov-Maxwell dispersion relation (dotted) and the estimated
frequency given by \eqref{eq:eq1} (dashed). Collisionless damping is
strong at $k_\parallel d_i$ values above the $\gamma/\omega = 0.2$
contour (thick line).}
\label{fig:kaw}
\end{figure}

In \figref{fig:kaw}, we plot contours of the value of $\log[\omega/(k
  v_A)]$ for the \Alfven mode determined using the linear
Vlasov-Maxwell dispersion relation (dotted) and the estimated
frequency given by \eqref{eq:eq1} (dashed). For a complex linear
eigenfrequency $\omega - i \gamma$, the linear collisionless damping
of the wave mode becomes strong at a value of $\gamma/\omega \gtrsim
0.2$.  Also plotted on \figref{fig:kaw} is the contour where
$\gamma/\omega = 0.2$ (thick solid); for values of $k_\parallel d_i$
above this line, the collisionless damping of the ion cyclotron wave
by the cyclotron resonance is sufficiently strong that the waves are
not expected to be observed in solar wind turbulence. For this
strongly damped region of wavevector space, strong wave-particle
interactions lead to deviations in the linear \Alfven mode frequency
from the expression in \eqref{eq:eq1}. In the weakly damped region
below the $\gamma/\omega = 0.2$, we see that the agreement between the
linear Vlasov-Maxwell real frequency and the estimate by
\eqref{eq:eq1} is excellent for both the \Alfven wave and kinetic
\Alfven wave regimes.

\begin{figure}[t]
\resizebox{3.1in}{!}{\includegraphics{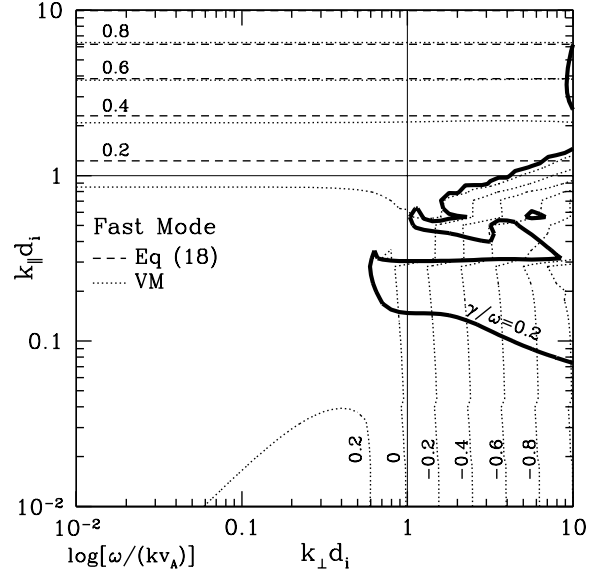}}
\caption{Contour plot of  $\log[\omega/(k v_A)]$  for the fast mode over the 
$(k_\perp d_i, k_\parallel d_i)$ plane for a $\beta_i=1$ and $T_i/T_e=1$ plasma determined by the linear
Vlasov-Maxwell dispersion relation (dotted) and the estimated
frequency given by \eqref{eq:whistler} (dashed). Collisionless damping is
strong at $k_\parallel d_i$ values above the $\gamma/\omega = 0.2$
contour (thick line).
}
\label{fig:whi}
\end{figure}

In \figref{fig:whi}, we plot contours of the value of $\log[\omega/(k
v_A)]$ for the fast mode determined using the linear Vlasov-Maxwell
dispersion relation (dotted) and the estimated frequency given by
\eqref{eq:whistler} (dashed). Again, linear collisionless damping
is strong for modes to the right (higher values of $k_\perp d_i$) of
the $\gamma/\omega = 0.2$ contour (thick solid). In the whistler wave
regime at $k_\parallel d_i \gtrsim 1$, agreement between the linear
Vlasov-Maxwell real frequency and the estimate by \eqref{eq:whistler}
is excellent. Note that the wave frequency in the ion Bernstein wave
regime ($k_\perp d_i \gtrsim 1$ and $k_\parallel d_i \ll 1$, see
\figref{fig:diagram}) is not well estimated by \eqref{eq:whistler}.

\section{Discussion}
\label{sec:discuss}

Here we discuss the implications of the results in \secref{sec:valid}
for the study of the dissipation range of solar wind turbulence.  The
primary result of this work is that the whistler wave is the only
linear kinetic wave mode that is likely to violate significantly the
Taylor hypothesis. This finding relies on the premise that the
frequency of the turbulent fluctuations is well characterized by the
frequency of the linear waves supported by the solar wind plasma. This
premise is contained within a more general hypothesis for the modeling
of plasma turbulence, the quasilinear premise; a detailed discussion
of the quasilinear premise and a review of supporting evidence is
presented in \citet{Howes:2014b}.  Adopting this premise, a key
question naturally arises about the nature of the dissipation range of
solar wind turbulence: do the turbulent fluctuations correspond to a
broadband spectrum of kinetic \Alfven waves, a broadband spectrum of
whistler waves, or some combination of both types of waves? Since we
predict that whistler waves will violate the Taylor hypothesis while
kinetic \Alfven waves will not, the answer to this question has
important implications for measurements of solar wind turbulence by
the upcoming \emph{Magnetospheric Multiscale (MMS)}, \emph{Solar
  Orbiter}, and \emph{Solar Probe Plus} missions.

The body of observational evidence in support of kinetic \Alfven waves
as the dominant contributor to the dissipation range of solar wind
turbulence has recently been reviewed in detail by
\citet{Podesta:2013}. Important lines of evidence in support of
kinetic \Alfven waves being the dominant contributor to turbulent
fluctuations in the dissipation range include enhanced electron
density fluctuations around the ion Larmor radius scale
\citep{Harmon:1989,Hollweg:1999,Chandran:2009b,Chen:2013b},
measurements of the wave frequency as a function of wavenumber
\citep{Bale:2005,Howes:2008b,Sahraoui:2010b,Salem:2012}, magnetic
helicity measurements around the ion Larmor radius scale
\citep{Howes:2010a,He:2011,Podesta:2011a,Klein:2014a}, the magnetic
field variance anisotropy at ion kinetic length scales
\citep{Belcher:1971,Harmon:1989,Leamon:1998a,Hollweg:1999,Smith:2006,Hamilton:2008,Gary:2009,TenBarge:2012b,Podesta:2012},
and the lack of compressible fast-wave fluctuations in the inertial
range of solar wind turbulence \citep{Howes:2012a,Klein:2012}.

In addition to the observational evidence above, there is also
evidence from numerical simulations that the solar wind dissipation
range consists of a cascade of kinetic \Alfven waves, including
gyrokinetic
\citep{Howes:2008a,Howes:2011a,TenBarge:2013a,TenBarge:2013b} and
electron reduced MHD simulations \citep{Boldyrev:2012b}, both of which
reproduce quantitatively the observed magnetic energy spectrum in the
dissipation range of the solar wind.

Although, based on the numerous lines of reasoning above, the
turbulent cascade of energy from large to small scales appears to be
dominated by kinetic \Alfven wave fluctuations, there is strong
evidence that kinetic instabilities may generate whistler waves at
scales $k d_i \sim 1$ \citep{Kasper:2002,Hellinger:2006,Bale:2009}.
Measurements of magnetic helicity sorted as a function of the angle of
the magnetic field with respect to the solar wind flow show a
subdominant contribution of opposite magnetic helicity at small
angles. These parallel modes are interpreted to be either whistler or
ion cyclotron waves with nearly parallel wavevectors
\citep{He:2011,Podesta:2011a,Podesta:2011b,Klein:2014a}.  Significant
energy input into the turbulence at scales $k d_i \sim 1$, however,
appears unlikely since structure in the turbulent energy spectra at
these scales is not generally observed (see \citet{Podesta:2009a} and
\citet{Wicks:2010} for exceptions, where a small bump can be seen in
the energetically subdominant $k_\parallel$ spectrum).  In addition,
modeling of the magnetic helicity as a function of angle using the
synthetic spacecraft data method shows that these parallel modes
contribute only around 5\% of the turbulent power, while the dominant
95\% contribution to the turbulent power is due to a spectrum of kinetic \Alfven waves with
nearly perpendicular wavevectors \citep{Klein:2014a}.

In conclusion, the lack of strong evidence for whistler waves in the
dissipation range is good news for planned single-spacecraft missions,
such as \emph{Solar Orbiter} and \emph{Solar Probe Plus}: we predict
that measurements of a kinetic \Alfven wave dominated dissipation
range will not violate the Taylor hypothesis, dramatically simplifying
data analysis and interpretation of turbulence measurements for these
upcoming missions.

In addition to the magnetic field measurements of turbulence
traditionally collected by spacecraft missions in the solar wind, many
new spacecraft missions are instrumented to make measurements of the
fluctuating turbulent electric fields as well. Relevant to this study
of the effect of measurements made in a frame of reference moving with
respect to the plasma being measured, it is important to note that one
must carefully handle the Lorentz transform of the electric fields, as
detailed in Appendix~\ref{app:lorentz}. The upshot is that, although
the magnetic fields may be safely transformed between spacecraft and
plasma frames without any complications, the Lorentz transform
relating the spacecraft-frame electric field $\V{E}_{sc}$ and the
plasma-frame electric field $\V{E}$ for typical solar wind conditions
is
\begin{equation}
\V{E}_{sc} =  \V{E} + \V{v}_{sw}/c \times \V{B}.
\label{eq:lorentz}
\end{equation}
The impact of this transformation is made clear in recent work
exploring the electric and magnetic field spectra using \emph{Cluster}
measurements \citep{Chen:2011a}, finding that previous electric field
spectra reported in the literature \citep{Bale:2005} were dominated by
the magnetic field spectrum through the second term in
\eqref{eq:lorentz}.

Finally, we discuss an important point that has lead to continuing
confusion in the literature regarding the relation between the ion
cyclotron frequency and spacecraft-frame measurements. The ion
cyclotron frequency $\Omega_i$ is a characteristic frequency of the
plasma \emph{in the plasma frame}---it therefore enters the spacecraft
frame frequency through the first term on the right hand side of
\eqref{eq:doppler}.  Therefore, purely temporal variation at the ion
cyclotron frequency in the plasma frame does not Doppler shift when
measured by a probe moving with respect to the plasma. Only spatial
variation Doppler shifts to yield a temporal fluctuation when measured
by a moving probe. Under typical solar wind conditions, the
spacecraft-frame frequency $\omega_{sc}$ of the break in the magnetic
energy frequency spectrum at the onset of the dissipation range is
often roughly coincident with the value of the ion cyclotron frequency
$\Omega_i$ in the plasma frame.  This fact has lead numerous
researchers to attribute the steepening of the spectrum to ion
cyclotron damping
\citep{Coleman:1968,Denskat:1983,Goldstein:1994,Leamon:1998a,Gary:1999,Hamilton:2008}.
But, unless the plasma-frame frequency term competes with, or
dominates, the Doppler shift term, $|\omega| \gtrsim |\V{k}\cdot
\V{v}_{sw}|$ (thereby significantly violating the Taylor hypothesis),
then any measurements of spacecraft-frame frequency at the ion
cyclotron frequency, $\omega_{sc}\sim \Omega_i$, are caused by the
condition $\V{k}\cdot \V{v}_{sw}\sim \Omega_i$, and are unrelated to
ion cyclotron frequency dynamics in the solar wind plasma frame,
$\omega \sim \Omega_i$. We have shown in \secref{sec:icw} that the
Taylor hypothesis is not generally violated for ion cyclotron waves.
Therefore, it is physically incorrect to interpret solar wind
turbulent fluctuations that have a spacecraft-frame frequency
$\omega_{sc}\sim \Omega_i$ as being related to ion cyclotron frequency
dynamics in the plasma frame.

\section{Conclusion}
\label{sec:conc}
The Taylor hypothesis is commonly used in the analysis of single-point
spacecraft measurements of solar wind turbulence to relate the
frequency of turbulent fluctuations measured in the spacecraft frame
directly to the spatial wavenumber of turbulent fluctuations in the
plasma frame. But, as upcoming missions, such as \emph{Solar Orbiter}
and \emph{Solar Probe Plus}, explore new observational regimes, the
Taylor hypothesis may fail. Two new regimes threaten to violate the
Taylor hypothesis: (i) Slow Flow Regime: when the solar wind flow velocity
falls below the \Alfven wave velocity, and (ii) Dispersive Regime:
when dispersive effects at small scales cause the wave frequency to
increase more rapidly than linearly with the wavevector.

To evaluate the validity of the Taylor hypothesis in these new
regimes, we adopt the premise that the frequency of the turbulent
fluctuations is well characterized by the frequency of the linear
waves supported by the solar wind plasma.  Therefore, we examine the
validity of the Taylor hypothesis for the linear kinetic wave modes in
the weakly collisional solar wind plasma. In particular, we focus on
two leading candidate wave modes: (i) the \Alfven wave and its
small-scale dispersive extension as the kinetic \Alfven wave, and (ii)
the kinetic counterpart of the MHD fast wave and its small-scale
dispersive extension as the whistler wave.

We present useful analytical expressions for the kinetic wave modes of
interest and numerically verify that these expressions are accurate by
direct comparison to solutions for the wave frequencies from the
linear Vlasov-Maxwell dispersion relation. We use those analytical
expressions to derive simple conditions for the validity of the Taylor
hypothesis.  Our principle finding is that \emph{the whistler wave is
  the only wave mode that is likely to violate significantly the
  Taylor hypothesis} when upcoming missions measure the turbulence in
the solar wind.  We predict that \emph{the Taylor hypothesis is not
  likely to be violated significantly by any of the other plasma
  waves} that may be relevant to turbulence in the solar wind: any
limit of the \Alfven mode including kinetic \Alfven waves and ion
cyclotron waves, the kinetic fast wave, or the kinetic slow wave.

We emphasize the importance of making the proper Lorentz
transformation of electric field measurements to relate the
measurements of the spacecraft-frame electric field to the
plasma-frame electric field, and we present the appropriate formula
for typical solar wind conditions. Finally, we demonstrate that it is
physically incorrect to interpret solar wind turbulent fluctuations
that have a spacecraft-frame frequency $\omega_{sc}\sim \Omega_i$ as
being related to ion cyclotron frequency dynamics in the plasma frame.

Significant evidence exists in support of the
kinetic-Alfv\'en-wave-like nature of the turbulent fluctuations in the
dissipation range. This is good news for upcoming single spacecraft
missions, such as \emph{Solar Orbiter} and \emph{Solar Probe Plus},
because it means that it will still be possible to apply the Taylor
hypothesis to \emph{in situ} measurements of the turbulence to relate
the spacecraft-frame frequency of the fluctuations to the spatial
wavenumber of fluctuations in the plasma frame, dramatically
simplifying the data analysis and interpretation for these upcoming
missions.

In a companion paper, \citet{Klein:2014b}, we employ the synthetic
spacecraft data method to explore the conditions leading to the
violation of the Taylor hypothesis and to predict the quantitative
effect on the measured magnetic energy spectrum in spacecraft-frame
frequency when the Taylor hypothesis is violated.

\acknowledgments

G.~G.~H. acknowledges Sylvia Perri, Foaud Sahraoui, and the rest of
ISSI Team 185 for inspiring discussions and helpful comments. Support
was provided by the International Space Science Institute in Bern, NSF
CAREER Award AGS-1054061, and NASA NNX10AC91G.

%

\appendix
\section{Lorentz Transform of  Electromagnetic Field Measurements in the Solar Wind}
\label{app:lorentz}
The general formula for the Lorentz transformation of electric and
magnetic fields from the unprimed frame $K$ (at rest) to the primed
frame $K'$ moving with velocity $\V{v}$ with respect to frame $K$ are
\begin{equation}
\V{E}' = \gamma ( \V{E} + \V{v}/c \times \V{B}) - \frac{\gamma^2}{\gamma+1} 
\V{v}/c (\V{v}/c \cdot \V{E}),
\label{eq:elor}
\end{equation}
\begin{equation}
\V{B}' = \gamma ( \V{B} - \V{v}/c \times \V{E}) - \frac{\gamma^2}{\gamma+1} 
\V{v}/c (\V{v}/c \cdot \V{B}),
\label{eq:blor}
\end{equation}
where $\gamma =1/\sqrt{1+v^2/c^2}$ \citep{Jackson:1998}.

For a velocity of the solar wind (the velocity of transformation between 
frames) $v\sim 500$~km/s and an \Alfven speed $v_A=50$~km/s, we have
$v/c=1.6\times 10^{-3}$ and $v_A/c=1.6\times 10^{-4}$, so $\gamma \simeq 1$.
For \Alfven waves, the characteristic wave electric fields are
\begin{equation}
\V{E} \sim \frac{v_A}{c} \delta \V{B} \times \zhat,
\end{equation}
so $\Order(E/\delta B) \sim v_A/c \ll 1$. Taking $v/c \sim \epsilon
\ll 1$, the order of the terms on the right-hand side of
\eqref{eq:elor} with respect to $\delta B$ is $\epsilon$, $\epsilon$,
and $\epsilon^2$.  So, the last term may be dropped, and we are left
with the approximate relation
\begin{equation}
\V{E}' \simeq  \V{E} + \V{v}/c \times \V{B}.
\end{equation}
The order of the terms on the right-hand side  of \eqref{eq:blor} with respect to
$\delta B$ is $1$, $\epsilon^2$, and $\epsilon^2$.  Therefore,
\eqref{eq:blor} reduces to 
\begin{equation}
\V{B}' = \V{B},
\end{equation}
and we do not need to concern ourselves with the Lorentz
transformation of the magnetic field from the spacecraft to the plasma
frame.


\begin{thebibliography}{74}
\expandafter\ifx\csname natexlab\endcsname\relax\def\natexlab#1{#1}\fi

\bibitem[{{Bale} {et~al.}(2009){Bale}, {Kasper}, {Howes}, {Quataert}, {Salem},
  \& {Sundkvist}}]{Bale:2009}
{Bale}, S.~D., {Kasper}, J.~C., {Howes}, G.~G., {Quataert}, E., {Salem}, C., \&
  {Sundkvist}, D. 2009, Phys.~Rev.~Lett., 103, 211101

\bibitem[{{Bale} {et~al.}(2005){Bale}, {Kellogg}, {Mozer}, {Horbury}, \&
  {Reme}}]{Bale:2005}
{Bale}, S.~D., {Kellogg}, P.~J., {Mozer}, F.~S., {Horbury}, T.~S., \& {Reme},
  H. 2005, Phys.~Rev.~Lett., 94, 215002

\bibitem[{{Barnes}(1966)}]{Barnes:1966}
{Barnes}, A. 1966, Phys.~Fluids, 9, 1483

\bibitem[{{Belcher} \& {Davis}(1971)}]{Belcher:1971}
{Belcher}, J.~W., \& {Davis}, L. 1971, J.~Geophys.~Res., 76, 3534

\bibitem[{{Boldyrev}(2006)}]{Boldyrev:2006}
{Boldyrev}, S. 2006, Phys.~Rev.~Lett., 96, 115002

\bibitem[{{Boldyrev} {et~al.}(2013){Boldyrev}, {Horaites}, {Xia}, \&
  {Perez}}]{Boldyrev:2013}
{Boldyrev}, S., {Horaites}, K., {Xia}, Q., \& {Perez}, J.~C. 2013,
  Astrophys.~J., 777, 41

\bibitem[{{Boldyrev} \& {Perez}(2012)}]{Boldyrev:2012b}
{Boldyrev}, S., \& {Perez}, J.~C. 2012, Astrophys.~J.~Lett., 758, L44

\bibitem[{{Bruno} \& {Carbone}(2005)}]{Bruno:2005}
{Bruno}, R., \& {Carbone}, V. 2005, Living Reviews in Solar Physics, 2, 4

\bibitem[{{Chandran} {et~al.}(2009){Chandran}, {Quataert}, {Howes}, {Xia}, \&
  {Pongkitiwanichakul}}]{Chandran:2009b}
{Chandran}, B.~D.~G., {Quataert}, E., {Howes}, G.~G., {Xia}, Q., \&
  {Pongkitiwanichakul}, P. 2009, Astrophys.~J., 707, 1668

\bibitem[{{Chen} {et~al.}(2011){Chen}, {Bale}, {Salem}, \&
  {Mozer}}]{Chen:2011a}
{Chen}, C.~H.~K., {Bale}, S.~D., {Salem}, C., \& {Mozer}, F.~S. 2011,
  Astrophys.~J.~Lett., 737, L41

\bibitem[{{Chen} {et~al.}(2013){Chen}, {Boldyrev}, {Xia}, \&
  {Perez}}]{Chen:2013b}
{Chen}, C.~H.~K., {Boldyrev}, S., {Xia}, Q., \& {Perez}, J.~C. 2013,
  Phys.~Rev.~Lett., 110, 225002

\bibitem[{{Cho} \& {Lazarian}(2003)}]{Cho:2003}
{Cho}, J., \& {Lazarian}, A. 2003, Mon.~Not.~Roy.~Astron.~Soc., 345, 325

\bibitem[{{Cho} \& {Lazarian}(2004)}]{Cho:2004}
---. 2004, Astrophys.~J.~Lett., 615, L41

\bibitem[{Cho \& Vishniac(2000)}]{Cho:2000}
Cho, J., \& Vishniac, E.~T. 2000, Astrophys.~J., 539, 273

\bibitem[{{Coleman}(1968)}]{Coleman:1968}
{Coleman}, Jr., P.~J. 1968, Astrophys.~J., 153, 371

\bibitem[{{Denskat} {et~al.}(1983){Denskat}, {Beinroth}, \&
  {Neubauer}}]{Denskat:1983}
{Denskat}, K.~U., {Beinroth}, H.~J., \& {Neubauer}, F.~M. 1983,
  J.~Geophys.~Zeit.~Geophys., 54, 60

\bibitem[{{Gary}(1999)}]{Gary:1999}
{Gary}, S.~P. 1999, J.~Geophys.~Res., 104, 6759

\bibitem[{{Gary} {et~al.}(2012){Gary}, {Chang}, \& {Wang}}]{Gary:2012}
{Gary}, S.~P., {Chang}, O., \& {Wang}, J. 2012, Astrophys.~J., 755, 142

\bibitem[{{Gary} \& {Smith}(2009)}]{Gary:2009}
{Gary}, S.~P., \& {Smith}, C.~W. 2009, J.~Geophys.~Res., 114, 12105

\bibitem[{Goldreich \& Sridhar(1995)}]{Goldreich:1995}
Goldreich, P., \& Sridhar, S. 1995, Astrophys.~J., 438, 763

\bibitem[{{Goldstein} {et~al.}(1994){Goldstein}, {Roberts}, \&
  {Fitch}}]{Goldstein:1994}
{Goldstein}, M.~L., {Roberts}, D.~A., \& {Fitch}, C.~A. 1994, J.~Geophys.~Res.,
  99, 11519

\bibitem[{{Hamilton} {et~al.}(2008){Hamilton}, {Smith}, {Vasquez}, \&
  {Leamon}}]{Hamilton:2008}
{Hamilton}, K., {Smith}, C.~W., {Vasquez}, B.~J., \& {Leamon}, R.~J. 2008,
  J.~Geophys.~Res., 113, A01106

\bibitem[{{Harmon}(1989)}]{Harmon:1989}
{Harmon}, J.~K. 1989, J.~Geophys.~Res., 94, 15399

\bibitem[{{He} {et~al.}(2011){He}, {Marsch}, {Tu}, {Yao}, \& {Tian}}]{He:2011}
{He}, J., {Marsch}, E., {Tu}, C., {Yao}, S., \& {Tian}, H. 2011, Astrophys.~J.,
  731, 85

\bibitem[{{Hellinger} {et~al.}(2006){Hellinger}, {Tr{\'a}vn{\'{\i}}{\v c}ek},
  {Kasper}, \& {Lazarus}}]{Hellinger:2006}
{Hellinger}, P., {Tr{\'a}vn{\'{\i}}{\v c}ek}, P., {Kasper}, J.~C., \&
  {Lazarus}, A.~J. 2006, Geophys.~Res.~Lett., 33, 9101

\bibitem[{{Hirose} {et~al.}(2004){Hirose}, {Ito}, {Mahajan}, \&
  {Ohsaki}}]{Hirose:2004}
{Hirose}, A., {Ito}, A., {Mahajan}, S.~M., \& {Ohsaki}, S. 2004, Physics
  Letters A, 330, 474

\bibitem[{{Hollweg}(1999)}]{Hollweg:1999}
{Hollweg}, J.~V. 1999, J.~Geophys.~Res., 104, 14811

\bibitem[{{Howes}(2008)}]{Howes:2008c}
{Howes}, G.~G. 2008, Phys.~Plasmas, 15, 055904

\bibitem[{Howes(2009)}]{Howes:2009b}
Howes, G.~G. 2009, Nonlin.~Proc.~Geophys., 16, 219

\bibitem[{{Howes}(2011)}]{Howes:2011c}
{Howes}, G.~G. 2011, Astrophys.~J., 738, 40

\bibitem[{{Howes} {et~al.}(2012){Howes}, {Bale}, {Klein}, {Chen}, {Salem}, \&
  {TenBarge}}]{Howes:2012a}
{Howes}, G.~G., {Bale}, S.~D., {Klein}, K.~G., {Chen}, C.~H.~K., {Salem},
  C.~S., \& {TenBarge}, J.~M. 2012, Astrophys.~J.~Lett., 753, L19

\bibitem[{{Howes} {et~al.}(2006){Howes}, {Cowley}, {Dorland}, {Hammett},
  {Quataert}, \& {Schekochihin}}]{Howes:2006}
{Howes}, G.~G., {Cowley}, S.~C., {Dorland}, W., {Hammett}, G.~W., {Quataert},
  E., \& {Schekochihin}, A.~A. 2006, Astrophys.~J., 651, 590

\bibitem[{{Howes} {et~al.}(2008{\natexlab{a}}){Howes}, {Cowley}, {Dorland},
  {Hammett}, {Quataert}, \& {Schekochihin}}]{Howes:2008b}
---. 2008{\natexlab{a}}, J.~Geophys.~Res., 113, A05103

\bibitem[{{Howes} {et~al.}(2008{\natexlab{b}}){Howes}, {Dorland}, {Cowley},
  {Hammett}, {Quataert}, {Schekochihin}, \& {Tatsuno}}]{Howes:2008a}
{Howes}, G.~G., {Dorland}, W., {Cowley}, S.~C., {Hammett}, G.~W., {Quataert},
  E., {Schekochihin}, A.~A., \& {Tatsuno}, T. 2008{\natexlab{b}},
  Phys.~Rev.~Lett., 100, 065004

\bibitem[{{Howes} {et~al.}(2014){Howes}, {Klein}, \& {TenBarge}}]{Howes:2014b}
{Howes}, G.~G., {Klein}, K.~G., \& {TenBarge}, J.~M. 2014, ArXiv e-prints,
  1404.2913

\bibitem[{{Howes} \& {Quataert}(2010)}]{Howes:2010a}
{Howes}, G.~G., \& {Quataert}, E. 2010, Astrophys.~J.~Lett., 709, L49

\bibitem[{Howes {et~al.}(2011)Howes, TenBarge, Dorland, Quataert, Schekochihin,
  Numata, \& Tatsuno}]{Howes:2011a}
Howes, G.~G., TenBarge, J.~M., Dorland, W., Quataert, E., Schekochihin, A.~A.,
  Numata, R., \& Tatsuno, T. 2011, Phys.~Rev.~Lett., 107, 035004

\bibitem[{{Ito} {et~al.}(2004){Ito}, {Hirose}, {Mahajan}, \&
  {Ohsaki}}]{Ito:2004}
{Ito}, A., {Hirose}, A., {Mahajan}, S.~M., \& {Ohsaki}, S. 2004, Phys.~Plasmas,
  11, 5643

\bibitem[{{Jackson}(1998)}]{Jackson:1998}
{Jackson}, J.~D. 1998, {Classical Electrodynamics, 3rd Edition} (New York:
  Wiley)

\bibitem[{{Kasper} {et~al.}(2002){Kasper}, {Lazarus}, \& {Gary}}]{Kasper:2002}
{Kasper}, J.~C., {Lazarus}, A.~J., \& {Gary}, S.~P. 2002, Geophys.~Res.~Lett.,
  29, 20

\bibitem[{{Klein} {et~al.}(2014{\natexlab{a}}){Klein}, {Howes}, \&
  {TenBarge}}]{Klein:2014b}
{Klein}, K.~G., {Howes}, G.~G., \& {TenBarge}, J.~M. 2014{\natexlab{a}},
  Phys.~Rev.~Lett., submitted

\bibitem[{{Klein} {et~al.}(2012){Klein}, {Howes}, {TenBarge}, {Bale}, {Chen},
  \& {Salem}}]{Klein:2012}
{Klein}, K.~G., {Howes}, G.~G., {TenBarge}, J.~M., {Bale}, S.~D., {Chen},
  C.~H.~K., \& {Salem}, C.~S. 2012, Astrophys.~J., 755, 159

\bibitem[{{Klein} {et~al.}(2014{\natexlab{b}}){Klein}, {Howes}, {TenBarge}, \&
  {Podesta}}]{Klein:2014a}
{Klein}, K.~G., {Howes}, G.~G., {TenBarge}, J.~M., \& {Podesta}, J.~J.
  2014{\natexlab{b}}, Astrophys.~J., 785, 138

\bibitem[{{Krauss-Varban} {et~al.}(1994){Krauss-Varban}, {Omidi}, \&
  {Quest}}]{Krauss-Varban:1994}
{Krauss-Varban}, D., {Omidi}, N., \& {Quest}, K.~B. 1994, J.~Geophys.~Res., 99,
  5987

\bibitem[{Leamon {et~al.}(1998)Leamon, Smith, Ness, Matthaeus, \&
  Wong}]{Leamon:1998a}
Leamon, R.~J., Smith, C.~W., Ness, N.~F., Matthaeus, W.~H., \& Wong, H.~K.
  1998, J.~Geophys.~Res., 103, 4775

\bibitem[{Maron \& Goldreich(2001)}]{Maron:2001}
Maron, J., \& Goldreich, P. 2001, Astrophys.~J., 554, 1175

\bibitem[{{Matthaeus} \& {Goldstein}(1982)}]{Matthaeus:1982b}
{Matthaeus}, W.~H., \& {Goldstein}, M.~L. 1982, J.~Geophys.~Res., 87, 6011

\bibitem[{{Narita} {et~al.}(2011){Narita}, {Gary}, {Saito}, {Glassmeier}, \&
  {Motschmann}}]{Narita:2011}
{Narita}, Y., {Gary}, S.~P., {Saito}, S., {Glassmeier}, K.-H., \& {Motschmann},
  U. 2011, Geophys.~Res.~Lett., 38, L05101

\bibitem[{{Perri} \& {Balogh}(2010)}]{Perri:2010a}
{Perri}, S., \& {Balogh}, A. 2010, Astrophys.~J., 714, 937

\bibitem[{{Podesta}(2009)}]{Podesta:2009a}
{Podesta}, J.~J. 2009, Astrophys.~J., 698, 986

\bibitem[{{Podesta}(2013)}]{Podesta:2013}
---. 2013, sp, 286, 529

\bibitem[{{Podesta} \& {Gary}(2011{\natexlab{a}})}]{Podesta:2011b}
{Podesta}, J.~J., \& {Gary}, S.~P. 2011{\natexlab{a}}, Astrophys.~J., 742, 41

\bibitem[{{Podesta} \& {Gary}(2011{\natexlab{b}})}]{Podesta:2011a}
---. 2011{\natexlab{b}}, Astrophys.~J., 734, 15

\bibitem[{{Podesta} \& {TenBarge}(2012)}]{Podesta:2012}
{Podesta}, J.~J., \& {TenBarge}, J.~M. 2012, Journal of Geophysical Research
  (Space Physics), 117, 10106

\bibitem[{{Quataert}(1998)}]{Quataert:1998}
{Quataert}, E. 1998, Astrophys.~J., 500, 978

\bibitem[{{Roberts} {et~al.}(2013){Roberts}, {Li}, \& {Li}}]{Roberts:2013}
{Roberts}, O.~W., {Li}, X., \& {Li}, B. 2013, Astrophys.~J., 769, 58

\bibitem[{{Sahraoui} {et~al.}(2012){Sahraoui}, {Belmont}, \&
  {Goldstein}}]{Sahraoui:2012}
{Sahraoui}, F., {Belmont}, G., \& {Goldstein}, M.~L. 2012, Astrophys.~J., 748,
  100

\bibitem[{{Sahraoui} {et~al.}(2010){Sahraoui}, {Goldstein}, {Belmont}, {Canu},
  \& {Rezeau}}]{Sahraoui:2010b}
{Sahraoui}, F., {Goldstein}, M.~L., {Belmont}, G., {Canu}, P., \& {Rezeau}, L.
  2010, Phys.~Rev.~Lett., 105, 131101

\bibitem[{{Saito} {et~al.}(2008){Saito}, {Gary}, {Li}, \&
  {Narita}}]{Saito:2008}
{Saito}, S., {Gary}, S.~P., {Li}, H., \& {Narita}, Y. 2008, Phys.~Plasmas, 15,
  102305

\bibitem[{{Salem} {et~al.}(2012){Salem}, {Howes}, {Sundkvist}, {Bale},
  {Chaston}, {Chen}, \& {Mozer}}]{Salem:2012}
{Salem}, C.~S., {Howes}, G.~G., {Sundkvist}, D., {Bale}, S.~D., {Chaston},
  C.~C., {Chen}, C.~H.~K., \& {Mozer}, F.~S. 2012, Astrophys.~J.~Lett., 745, L9

\bibitem[{{Schekochihin} {et~al.}(2009){Schekochihin}, {Cowley}, {Dorland},
  {Hammett}, {Howes}, {Quataert}, \& {Tatsuno}}]{Schekochihin:2009}
{Schekochihin}, A.~A., {Cowley}, S.~C., {Dorland}, W., {Hammett}, G.~W.,
  {Howes}, G.~G., {Quataert}, E., \& {Tatsuno}, T. 2009, Astrophys.~J.~Supp.,
  182, 310

\bibitem[{Shebalin {et~al.}(1983)Shebalin, Matthaeus, \&
  Montgomery}]{Shebalin:1983}
Shebalin, J.~V., Matthaeus, W.~H., \& Montgomery, D. 1983, J.~Plasma Phys., 29,
  525

\bibitem[{{Smith} {et~al.}(2006){Smith}, {Hamilton}, {Vasquez}, \&
  {Leamon}}]{Smith:2006}
{Smith}, C.~W., {Hamilton}, K., {Vasquez}, B.~J., \& {Leamon}, R.~J. 2006,
  Astrophys.~J.~Lett., 645, L85

\bibitem[{{Smith} \& {Zhou}(2007)}]{Smith:2007}
{Smith}, E.~J., \& {Zhou}, X. 2007, in American Institute of Physics Conference
  Series, Vol. 932, Turbulence and Nonlinear Processes in Astrophysical
  Plasmas, ed. D.~{Shaikh} \& G.~P. {Zank}, 144--152

\bibitem[{{Stix}(1992)}]{Stix:1992}
{Stix}, T.~H. 1992, {Waves in Plasmas} (New York: American Institute of
  Physics)

\bibitem[{{Stringer}(1963)}]{Stringer:1963}
{Stringer}, T.~E. 1963, Journal of Nuclear Energy, 5, 89

\bibitem[{{Taylor}(1938)}]{Taylor:1938}
{Taylor}, G.~I. 1938, {Proc. Roy. Soc. A}, 164, 476

\bibitem[{{TenBarge} \& {Howes}(2012)}]{TenBarge:2012a}
{TenBarge}, J.~M., \& {Howes}, G.~G. 2012, Phys.~Plasmas, 19, 055901

\bibitem[{{TenBarge} \& {Howes}(2013)}]{TenBarge:2013a}
---. 2013, Astrophys.~J.~Lett., 771, L27

\bibitem[{{TenBarge} {et~al.}(2013){TenBarge}, {Howes}, \&
  {Dorland}}]{TenBarge:2013b}
{TenBarge}, J.~M., {Howes}, G.~G., \& {Dorland}, W. 2013, Astrophys.~J., 774,
  139

\bibitem[{{TenBarge} {et~al.}(2012){TenBarge}, {Podesta}, {Klein}, \&
  {Howes}}]{TenBarge:2012b}
{TenBarge}, J.~M., {Podesta}, J.~J., {Klein}, K.~G., \& {Howes}, G.~G. 2012,
  Astrophys.~J., 753, 107

\bibitem[{{Tu} \& {Marsch}(1995)}]{Tu:1995}
{Tu}, C.-Y., \& {Marsch}, E. 1995, Space Sci.~Rev., 73, 1

\bibitem[{{Wicks} {et~al.}(2010){Wicks}, {Horbury}, {Chen}, \&
  {Schekochihin}}]{Wicks:2010}
{Wicks}, R.~T., {Horbury}, T.~S., {Chen}, C.~H.~K., \& {Schekochihin}, A.~A.
  2010, Mon.~Not.~Roy.~Astron.~Soc., 407, L31

\bibitem[{{Yoon} \& {Fang}(2008)}]{Yoon:2008}
{Yoon}, P.~H., \& {Fang}, T.-M. 2008, Plasma Phys.~Con.~Fus., 50, 125002

\end{thebibliography}

\end{document}